\documentclass[12pt]{article}
\newfont{\feff}{cmti10}
\usepackage{graphicx}
\usepackage{subfigure}
\usepackage{epstopdf}

\begin{document}

\title{Probability Densities  in Strong Turbulence.}
\author{Victor Yakhot \\
Department of Aerospace and Mechanical Engineering,\\ Boston
University, Boston 02215}

\maketitle 
\begin{abstract}
\noindent  According to modern developments in turbulence theory,  the "dissipation" scales (u.v. cut-offs)  
$\eta$ form 
a random field related to  velocity increments  $\delta_{\eta}u$. In this work  we,
using Mellin's transform combined with the Gaussain  large -scale boundary  condition, 
 calculate  probability densities  (PDFs) of velocity increments $P(\delta_{r}u,r)$ and the PDF  of
the  dissipation scales $Q(\eta, Re)$, where $Re$ is the large-scale Reynolds number.    The resulting expressions  strongly  deviate from the Log-normal PDF  $P_{L}(\delta_{r}u,r)$ often quoted in the literature.  It is shown that  the   probability density of the small-scale velocity fluctuations  includes  information about the large (integral) scale dynamics  which is responsible for deviation of $P(\delta_{r}u,r)$ from $P_{L}(\delta_{r}u,r)$.   A  framework for evaluation of the PDFs of various turbulence characteristics involving spatial derivatives is developed.  The exact relation, free of spurious Logarithms  recently discussed in Frisch et al  (J. Fluid Mech. {\bf 542}, 97 (2005)),   for the multifractal probability density of velocity increments, not based on the steepest descent evaluation of the integrals is obtained and the calculated function $D(h)$ is close to experimental data.  A novel   derivation  (Polyakov,  2005),   of a well-known result of the multi-fractal theory [Frisch,  "Turbulence. {\it  Legacy of A.N.Kolmogorov}", Cambridge University Press, 1995)) ,   based on  the concepts  described  in this paper,  is also presented.
 \end{abstract}
 \newpage

\section{Introduction.}

\noindent    A reasonably  well  experimentally established  
anomalous (multi) scaling of  the structure functions  $S_{n,m}=\overline{(u(x+r)-u(x))^{n}(v(x+r)-v(x))^{n}}\equiv \overline{ (\delta_{r}u)^{m}(\delta_{r}v)^{m}}$ , where $u$ and $v$ are components of the velocity field parallel and perpendicular to the displacement vector ${\bf r}=r{\bf i}$, respectively,   is one of  the properties  of  strong turbulence which makes it "the last unsolved problem of continuum mechanics".  [1]-[4].   The  anomalous dimension,  a property  of strongly interacting systems first introduced to  the quantum field theory by Gribov and Migdal [5] and Polyakov [6],    is a notoriously difficult from theoretical viewpoint  concept and  only recently, after many years of trying,  the theory of anomalous scaling of the structure functions of a  passive scalar advected by  a white-in-time random velocity field, has been developed [7]-[8].  Almost simultaneously,    the theory of bi-scaling in  turbulence generated by  the forced  Burgers equation [9]-[10]  was formulated.  
\noindent The attempts to explain anomalous dimensions in three dimensional turbulence were made in Refs.[11]-[12].  It was shown that the  Navier-Stokes equations combined with a simple model for the pressure -velocity correlations , lead to  homogeneous differential equations for the structure functions and, as a result, to anomalous scaling exponents which cannot be obtained on dimensional grounds. 

\noindent  During last forty-fifty years, a  substantial effort was devoted to derivation or at least modeling of various  probability densities in turbulence. The first attempts  resulting in the Log-normal PDF  of the dissipation rate fluctuations, consistent with  the multi-scaling,   
were made  by Kolmogorov [13] and Yaglom [14], using a simple cascade model. This   result  was later criticized by Orszag [15] as,  in general, not   realizable.  Similar Log-normal PDF  was obtained for the  not too large magnitudes of  velocity increment $\delta_{r}u=u(x+r)-u(x)$ in Ref. [11] , which was later experimentally tested by Kurien and Sreenivasan [3].  Some other attempts based on  analysis of experimental data led to various fits ranging from Log-normal and  Log-Poison expressions to exponential and "stretched exponential "  PDFs  $P(\delta_{r}u)$.  In this paper we, addressing this problem, will 
restrict ourselves by  considerations based on equations of motion. 

\noindent This paper is organized in a following way. In the Section 2 we introduce the Mellin transform for the probability density of velocity increments  and define the large- scale  Gaussian boundary condition.  We  show that the "normal  (linear ) scaling" corresponds  to the Gaussaian PDF of the small-scale fluctuations.  The PDF accounting for small deviations from the linear scaling is calculated   and compared with traditional Log-normal expressions.  In Section 3, we based on  theoretical analysis of the Navier-Stokes equations and numerical simulations by Gotoh and Nakano [16],  justify the boundary conditions used for the derivation of Sec 2.  The random field of the  dissipation scales 
$\eta$ (linear dimensions of the dissipative structures) , derived  from  the expression for the dissipation anomaly,  is introduced in Sec 4 and the PDF $Q(\eta, Re)$ is computed in Section 5.  Discussions and conclusions are presented in Section 6. There, the exact  expression for the probability density of velocity increments  following the multifractal formalism is presented and one of the results of the multi-fractal theory is derived (Polyakov 2005) without the multifractal input.   Some of the concepts and relations  presented  below have been reported in the previous publications [12],  [17]-[18].  Wherever needed, we  repeat them here  for the sake of continuity and clarity. \\

\section{The Mellin transform.}

\noindent  If the moments of velocity increments $\delta_{r}u$ are given by the scaling relations $S_{n,0}=\overline{(\delta_{r}u)^{n}}=A(n)r^{\xi_{n}}$, 
then  
probability density function can be found from the  Mellin transform:

\begin{equation}
P(\delta_{r}u,r)=\frac{1}{\delta_{r} u}\int_{-i\infty}^{i\infty} A(n)r^{\xi(n)}(\delta_{r}u)^{-n}dn
\end{equation}

\noindent   where 
we set the integral scale $L$ 
and the dissipation  rate ${\cal E}$ equal to unity. Multiplying (1) by $(\delta_{r}u)^{k}$ and evaluating a simple integral, gives  $S_{k,0}=A(k)r^{\xi_{k}}$. 
 
\noindent To make use of the relation (1) the dynamic information about both the amplitudes $A(n)$  and the exponents $\xi_{n}$ is needed.   In order to obtain  an  expression for $A(n)$  from  a large scale boundary condition on the PDF,   we first  have to define the integral scale $L$.  Based on experimental  data and theoretical consideration (see below), we choose  $L$ as a scale at which the energy flux toward small scales changes  sign or tends to zero. This means that  at the small scales $r<L$ the structure function $S_{3}(r)<0$
and at the larger scales  $r>L$, $S_{3}\geq 0$. Typically,  at this scale which depends upon the geometric details of the flow, the odd-order moments $S_{2n+1}(L)=0$
and the even-order  ones saturate: $\partial _{r}S_{2n,0}(L)=0$.  This property of turbulence has recently been examined both numerically and experimentally [19]-[20] (a  theoretical argument  will be given below). The scale $L$ appears naturally in a simple case of the Navier-Stokes equations  on an infinite domain driven by the white-in-time forcing function with the variance: 

$$\overline{f^{2}(k)}=\frac{{\cal P}}{2(2\pi)^{4}}\delta(k-k_{f})/k^{2}$$

\noindent where ${\cal P}$ is the forcing power. The exact calculation of the relation for the third-order structure function $S_{3}(r)$ gives an oscillating expression

$$S_{3}=-{\cal P}\frac{-36r\cos r+12\sin r-12(-2+r^{2})\sin r}{r^{4}} $$

\noindent In the limit $r\rightarrow 0$, we have the Kolomogorov relation $S_{3}=-\frac{4}{5}{\cal P}r$. 
At the large scale   $L\approx 5.88/k_{f}$ , $S_{3}(L)= 0$ [21].  In all flows  studied, the probability density $P(\delta_{L}u,L)$ was extremely close to the  Gaussian.  We would like to stress that the integral scale defined this way is not the  largest scale (size)  a  system but rather   corresponds to the top of the inertial range where the constant  energy flux toward small scales sets up. 
The Gaussian boundary condition at $r=1$ leads to a plausible and well- tested  (both experimentally and numerically),  expression $A(n)=(2n-1)!!$ which will be used in all calculations of this paper.  In what  follows a  dynamic and numerical  justification for this result  will be presented. 

\noindent First we consider the  case of  "normal scaling " $\xi=an$. 
Writing $(2n-1)!!=\frac{2^{n}}{\sqrt{\pi}}\int_{-\infty}^{\infty}e^{-x^{2}}x^{2n}dx$ and rotating the integration axis by $90^{o}$, we have setting $\delta_{r}u\equiv u$:

\begin{equation}
P(u,r)=\frac{1}{\sqrt{\pi} u}\int_{-\infty}^{\infty}e^{-x^{2}}dx\int_{-\infty}^{\infty} e^{in( \ ln \frac{r^{a}\sqrt{2}}{u}+ ln\ x)}dn
=\frac{1}{\sqrt{\pi} u}\int_{-\infty}^{\infty}e^{-x^{2}}\delta( \ ln \frac{r^{a}\sqrt{2}}{u}+ ln\ x)dx
\end{equation}
\noindent This integral is evaluated readily with the result: 

$$P(u)=\frac{1}{\sqrt{2 \pi}r^{a}}e^{-(\frac{u^{2}}{2r^{2a}})}$$

\noindent Now we   consider anomalous scaling by introducing small deviations from  the linear relation: 
\begin{equation}
\xi(n)=an-bn^{2}
\end{equation}

\noindent  which for not too large moment numbers $n$  can perceived as first two terms of the  Taylor expansion  of $\xi_{n}$ in the vicinity of $n=0$.  The formula  (3) is a generic {\it perturbative}  expression for the exponents $\xi_{n}$ independent upon the nature of the problem and the relations  similar to (3)  have resulted from the recent   perturbative theories   of a passive scalar in  a random velocity field. [7]-[8].    
 Using the Kolmogorov constraint $\xi_{3}=1$,  gives:  $b=(3a-1)/9$. 
It is  clear that the expression  (3) cannot be correct for all values of the moment order $n$. Indeed, in accord with Holder's  inequality,  $\xi_{n}$, is a concave and non-decreasing function of $n$ or in other words as $n\rightarrow \infty$, $\xi(n)/n\rightarrow 0$ ~($\xi_{n}/n\gg 1/n$).
Still for $n\leq 10-15$, the experimental data on strong turbulence are consistent with 
$a\approx 0.383$ and $b\approx 0.0166$ and the expressions derived below can be accurate only for not-too-large values of  velocity increment $\delta_{r}u$.  Thus, the probability density is given by the integral:

\begin{equation}
P(u,r)=\frac{1}{\sqrt{\pi} u}\int_{-\infty}^{\infty}e^{-x^{2}} \int_{-\infty}^{\infty} 
e^{in \ ln \frac{r^{a}\sqrt{2}x}{u}-bn^{2}ln\ r}dn
\end{equation}

\noindent which is reduced to:
\begin{equation}
P(u,r)=\frac{2}{\pi u\sqrt{4 \ln r^{b}}}\int_{-\infty}^{\infty}e^{-x^{2}}
exp[-\frac{ (\ln \frac{u}{r^{a}\sqrt{2}x})^{2}}{4b\ln r}]dx
\end{equation}

\noindent  The integration over $n$ leading from (4) to (5) was based on the following estimate.
 Since $e^{in\ln(u/r^{a})}$ is an oscillating function, the main contribution to the integral over  $n$ comes from the interval $n\approx  1/\sqrt{b|\ln r|}$     and in order for the relation (3) to be valid  in the integration interval, the following condition must be satisfied: $1/\sqrt{b|\ln r|}\leq 10-15 $ which is satisfied when the displacement $r$ is small enough.

\noindent The PDF $P(\frac{u}{r^{a}},r)$  numerically evaluated from equation (5), is shown  on Fig. 1.  for a few values of displacement $r$. We can see that the tails of the PDF strongly depend upon $r$ which is  a sign of intermittency and anomalous scaling. 

\begin{figure}[h]
  \center
  \subfigure[]{\includegraphics[height=6cm]{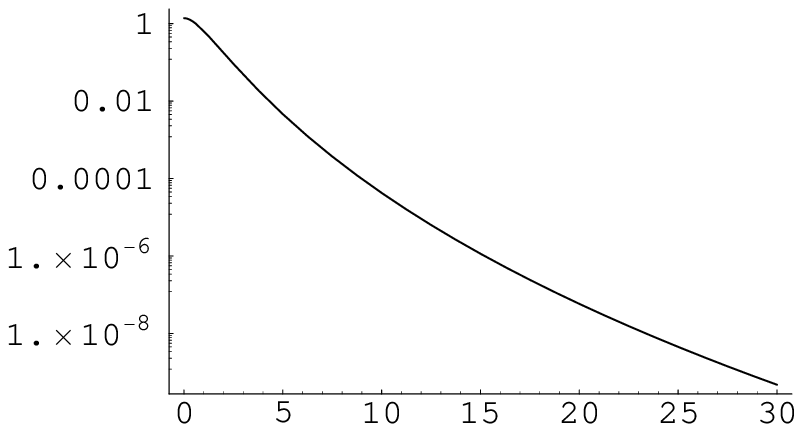}}
  \subfigure[]{\includegraphics[height=6cm]{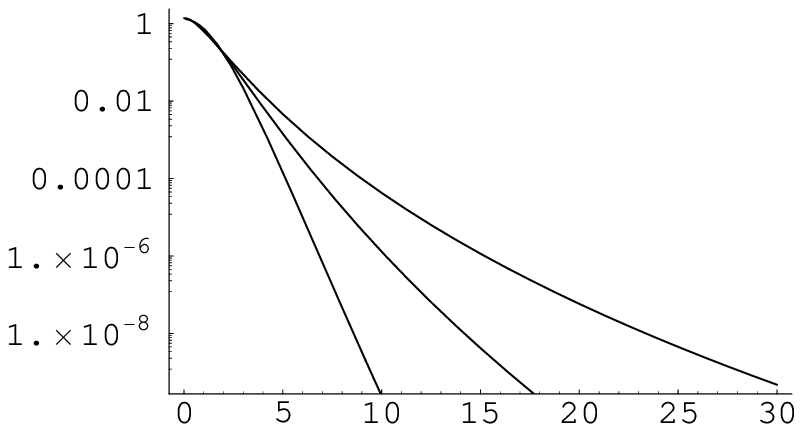}}
  \caption{Probability density $P(\frac{\delta_{r}u}{r^{a}},r)$ vs $\delta_{r}u/r^{a}$. a.~r=0.01; b~r=0.01;~0.1;~0.5}
\end{figure}

\noindent In the range $u/r^{a}\approx  1$,  the integral is dominated by  the interval $ x\approx 1$
leading to  the log-normal  result:  

\begin{equation}
P_{L}= \frac{1}{u\sqrt{4\pi b\ln r}}exp[-\frac{ln^{2}(\frac{r^{a}}{ u})}{4b \ln  r}] 
\end{equation}

\noindent  Both PDFs (5) and (6) are normalized to two. The  expression (6) , which  has  also been derived directly from the Navier-Stokes equations combined with a  simple model for the pressure -velocity correlation function in Ref. [11], has been  experimentally verified by Kurien and Sreenivasan [3].   The comparison between  (5) and (6)  showing a  surprisingly large  difference between the approximately  and numerically calculated integral (5)  is presented on Fig. 2.   In addition, it follows from  Figs. 1 and 2  that while $P(\delta_{r} u,r)$ has a maximum at $\delta_{r}u=0$,   in the limit $\delta_{r}u \rightarrow 0$,  the Log-normal PDF $ P_{L}$, 
given by (6) rapidly decreases to zero, meaning that it may be a reasonable  approximation not too close to the origin at $\delta_{r}u=0$.


  

\begin{figure}[h]
  \center
  \subfigure[]{\includegraphics[height=6cm]{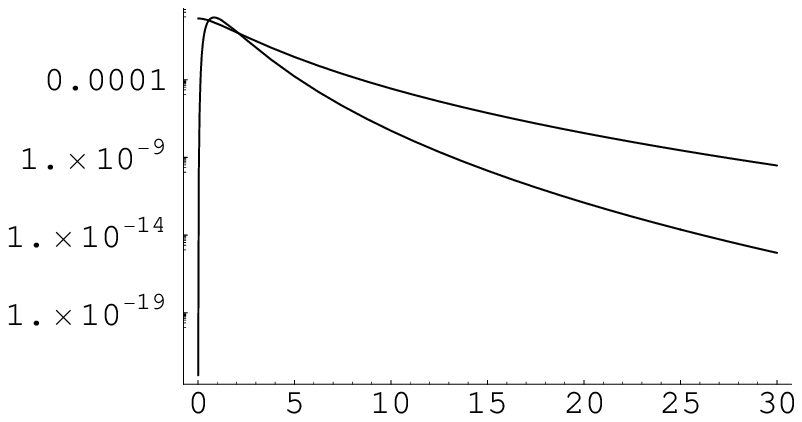}}
  \subfigure[]{\includegraphics[height=6cm]{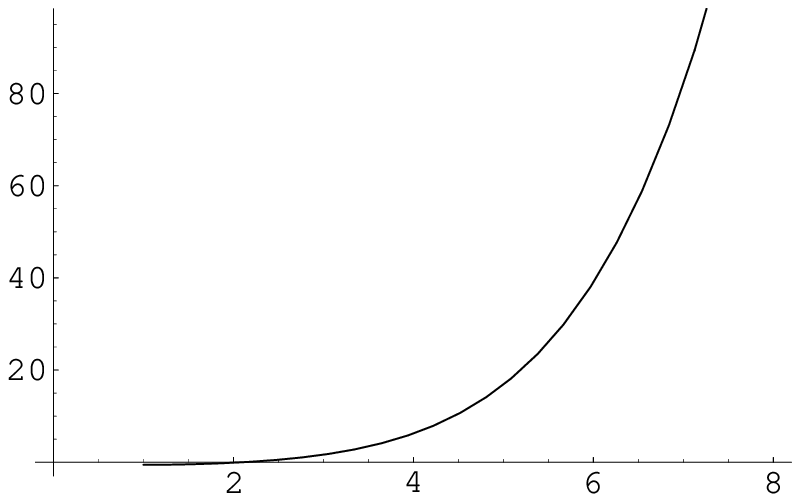}}
  \caption{a. ~Computed PDF $P(\frac{\delta_{r}u}{r^{a}},r)$ vs $\delta_{r}u/r^{a}$ (eq (5))  and Log-normal PDF  $P_{L}(\frac{\delta_{r}u}{r^{a}},r)$ 
  (eq. (6) ) for the same values of parameters.~b.~$(P-P_{L})/P_{L}$  vs $\delta_{r}u/r^{a}$.}
\end{figure}


\begin{figure}[h]
 \center
 \subfigure[] {\includegraphics[height=8cm]{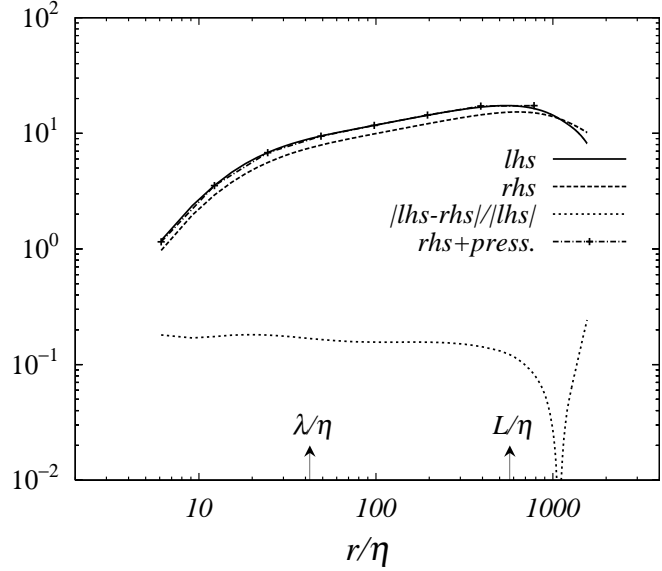}}
\subfigure[]{ \includegraphics[width=8cm]{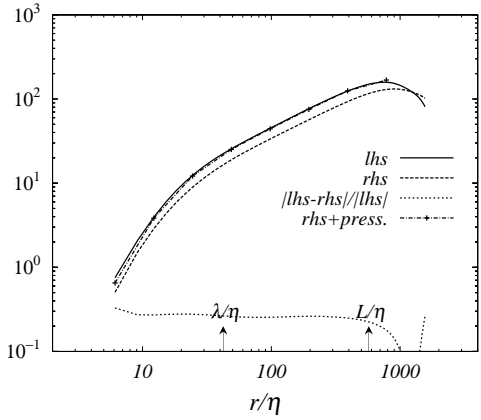}}
\subfigure[] {\includegraphics[width=8cm]{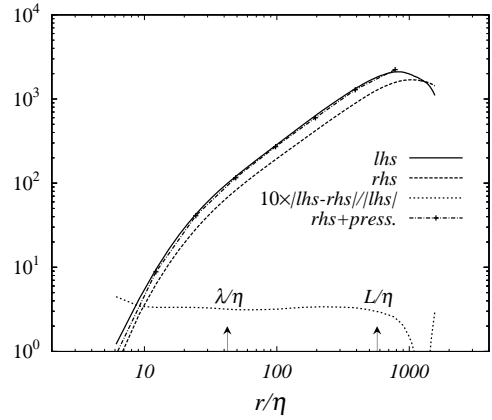}}
\caption{From Gotoh and Nakano (Ref.[16]). The top curves show the right  ("rhs") and left ("lhs") sides of equation (8) for $n=4;~6;~8$. When the pressure contribution was added to the "rhs", very accurate equality  $rhs=lhs$ has been reached (dotted line on top of the $"lhs"$  curve).  The bottom curves are $R_{2n}=-\frac{(2n-1)\overline{\delta_{r}(\partial_{x}p)u^{2n-2}}}{\partial_{r}S_{2n,0}+2S_{2n,0}/r}$ . The vertical axis of the Fig.3c is scaled by factor 10.}
\end{figure}

 


\begin{figure}[h]
  \center
  \subfigure[]{\includegraphics[height=6cm]{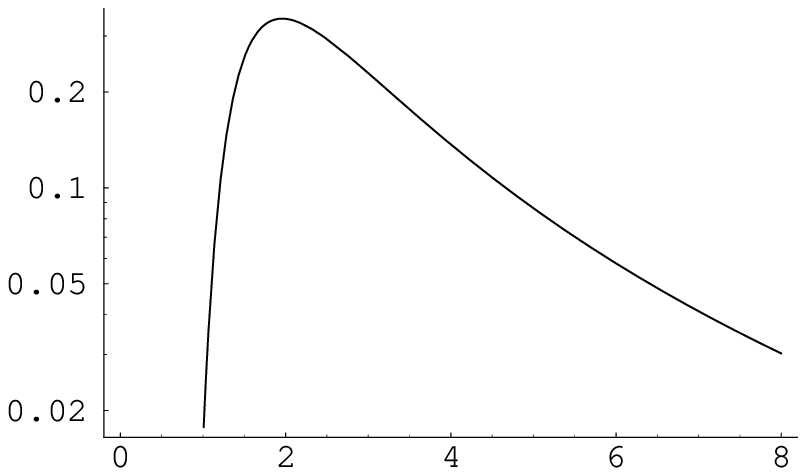}}
  \subfigure[]{\includegraphics[height=6cm]{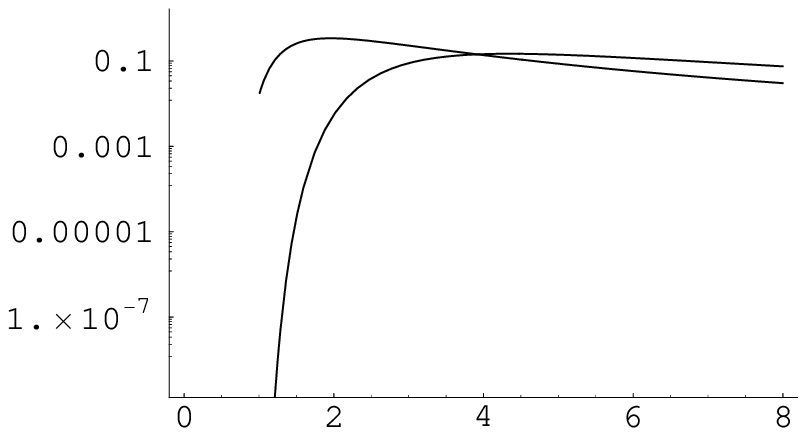}}
  \caption{Probability density $Q(\eta/\eta_{K},Re)$ vs $\eta/\eta_{K}$ where $\eta_{K}$ is the Kolmogorov scale. a. The shape of the PDF $Q(\eta,Re)$ for $Re\approx 0.3$.~b.  for $Re=0.1;~~0.3$.}
  \label{fig:enstrophy_decay}
\end{figure}



\noindent In the most interesting case $\xi(n)=\frac{an}{1+bn}$, the integral (1) with $A(n)=(2n-1)!!$ can be evaluated  both exactly and by the steepest descent method leading at the large magnitudes of the argument $u/r^{a}$ to the algebraically decreasing probability density consistent with the saturation 
of the exponents $\xi_{n}\rightarrow a/\beta$. \\

\section{The relations between moments. Simulations by Gotoh and Nakano (Ref. [16].)}

\noindent  In this Section  we would like  to develop some dynamic arguments justifying the choice of the structure functions amplitudes $A_{2n,0}=(2n-1)!! $ used in the above calculations.
\noindent   The relations  for the moments of velocity difference $S_{n,m}$ were derived in Refs. [11], [12]  and later in [22] using an alternative approach.
In particular,  the equation for the even-order moments is:

 \begin{equation}
 \frac{\partial S_{2n,0}}{\partial r}+\frac{d-1}{r}S_{2n,0}-\frac{(d-1)(2n-1)}{r}S_{2n-2,2}=(2n-1)<\delta_{r}  a u^{n-2}>
 \end{equation}
\noindent and taking into account that the dissipation contribution can be neglected  [11], [12], [3], [22]: we have

 \begin{equation}
 \frac{\partial S_{2n,0}}{\partial r}+\frac{d-1}{r}S_{2n,0}-\frac{(d-1)(2n-1)}{r}S_{2n-2,2}=-(2n-1)<\delta_{r} (\partial_{x}p) u^{2n-2}>
 \end{equation}

\noindent This relation is not closed. In the vicinity of the integral scale where the odd-order moments $S_{2n+1}(L)$ and derivatives of the even order moments  $\partial_{r}S_{2n,0}(L)$ disappear, the pressure gradient terms can be neglected. Here we would like to present a possible  physical explanation of  this effect.  If geometry  of  the large- scale unstable structures leading to the energy cascade toward small scales is  defined by the condition $S_{3}(L(x,y,z))=0$, then associating the instability with a flow separation phenomenon, gives the standard relation (see any textbook on hydrodynamics): 
$\nabla p=\nu\nabla^{2}{\bf u}$,  meaning that at these points  $a=0$ and $\delta_{L}{\bf a}=0$.  
This mechanism, if locally  correct allows one to neglect  acceleration contributions to  the equation (7) (or pressure gradients  in (8)) with the remaining terms giving  $S_{2n,0}=(2n-1)S_{2n-2,2}$, 
consistent with the Gaussian distribution.  Thus, extrapolating this result into the "inertial range", we assume that $(2n-1) S_{2n-2,2}=S_{2n,0}S_{0,2}/S_{2,0}=(1+\frac{\xi_{2}}{2})S_{2n,0}$ which is consistent with the Gaussian distribution at $r=1$.   This relation can be proven as follows:  since $\overline{uv}=0$, at a gaussian point $r=1$ we have $S_{2n-2,2}=S_{0,2}S_{2n-2,0}=\frac{1}{2n-1}\frac{S_{0,2} }{S_{2,0}}S_{2n,0}=\frac{1+\xi_{2}/2}{2n-1}S_{2n.0}$.  One remark is in order:  in the inertial range $\xi_{2}\neq 0$ while in the limit $r \rightarrow L$, the $r$-derivatives disappear  or in other words loosley defined in this limit "exponent"  $\xi_{2} \rightarrow 0$. Thus the above relation has to be treated as a parametrization valid in both limits $r\rightarrow L$ and $r/L\ll 1$.
In their remarkable paper   Gotoh and Nakano [16] made a detailed examination of the various terms in the equations (7) and (8).  Some of their results are presented on Fig. 3. where the left side of equations (7)-(8) denoted as "lhs" is compared with the right side ("rhs"minus pressure-gradient terms) for $n=4;~6;~8$. On the Fig. 3 $\lambda$ stands for the Taylor micro-scale. 
They demonstrated  that : 1.~ indeed , the dissipation contributions to the equations for the even order moments are negligibly small;  2. ~the relation $(2n-1)S_{2n-2,2}\approx 1.35 S_{2n,0}$,  independently upon the displacement magnitude $r$  (not shown on Fig. 3); 3.~The pressure contributions $(2n-1)\overline{\delta_{r}(\partial_{x}p)( \delta_{r}u )^{2n-2}}/lhs\approx const$  in the entire range of the displacement $r$-variation.  
\noindent  Substituting $(2n-1)S_{2n-2,2} \approx 1.35 S_{2n,0}$  and the algebraic relation  $S_{2n,0}=A_{2n,0}r^{\xi_{2n}}$ into ( 8)  gives ($d=3$):

 \begin{equation}
R_{2n}= \frac{(\partial_{r}+\frac{2}{r})S_{2n,0}-\frac{2(2n-1)}{r}S_{2n-2,2}}{  (\partial_{r}+\frac{2}{r})S_{2n,0} }=\frac{\xi_{2n}-\xi_{2}}{\xi_{2n}+2}=-\frac{(2n-1)<\delta_{r} (\partial_{x}p) u^{2n-2}>}{\frac{\partial S_{2n,0}}{\partial r}+\frac{2}{r}S_{2n,0}}
 \end{equation}

 \noindent and  for $\xi_{2}\approx  0.7$, $\xi_{4}\approx 1.27$,  $\xi_{6}\approx 1.77$ and $\xi_{8}\approx 2.15$ the magnitudes of the ratio:  $R_{4}=0.18;~R_{6}=0.28$ and $R_{8}=0.35$ in a close agreement with numerical data of Ref. [16] (For comparison, see  Fig.3).
 Moreover, one can see form Fig. 3 that all  pressure-gradient velocity correlations become negligibly small in the vicinity of  a single length scale ($L\approx 1000\eta_{K})$  consistent with  the above introduced definition of integral scale $L$.   Here $\eta_{K}$ is the Kolmogorov scale. The equations  (8) combined with the above considerations yield another interesting  relation for the pressure gradient-velocity correlations:
 
 $$(\xi_{2n}-\xi_{2})S_{2n,0}= -(2n-1)r\overline{\delta_{r}(\partial_{x}p)(\delta_{r}u)^{2n-2}}\geq 0$$

 \noindent The most important outcome of the formula (9) and numerical data of Ref. [16] is that the ratios $R_{2n}=const$ are independent upon the magnitude of displacement $r$. This means that the Lagrangian acceleration contribution to (8)  can be expressed as a linear combination:
 
 \begin{equation}
-<\delta_{r}\partial_{x}p|\delta u,\delta v>= \alpha_{p}\partial_{r}(\delta_{r} u)^{2}+\beta_{p}(\delta_{r}u)^{2}/r+\gamma_{p}(\delta_{r}v)^{2}/r+\kappa_{p}\partial_{r}(\delta_{r} v)^{2}
 \end{equation}
 \noindent plus sub-leading terms which are irrelevant in the small-scale limit $r\rightarrow 0$. \\

\section{Dissipation anomaly. Calculation of spatial derivatives}

 \noindent Similar result is obtained if we consider the 
 {\it \   dissipation anomaly} introduced by Polyakov  [9] who was  interested in turbulence generated by the random-force-driven Burgers equation. Later Polyakov's results have been generalized to the Navier-Stokes equations by Duchon and Robert[24],  Eyink [24] and Yakhot and Sreenivasan [18].
Interested in various characteristics of the velocity field, one often needs to calculate spatial derivatives 
defined as usual: 

$$\partial_{x}u=\lim_{y\rightarrow 0}\frac{u(x+y)-u(x)}{y}$$

\noindent  The problem is that if the velocity field is differentiable, then $S_{3}=O(y^{3})$  and $\partial_{y}S_{3}\rightarrow 0$. This is in contradiction with the Kolmogorov relation stating $\partial_{y}S_{3}\propto {\cal E}=O(1)$. This means that in the "inertial range"  the velocity field is singular. To resolve this problem, Polyakov  [9] introduced a limit-ordering procedure: 1.~$\nu\rightarrow 0$;~2. $y \rightarrow 0$. Assuming that there exist a scale ("dissipation") separating analytic and singular ranges, we can  approximately redefine  derivative as [18]: 

\begin{equation}
\partial_{x} u(x)=\lim_{y\rightarrow\eta\rightarrow 0} (u(x+y)-u(x-y))/2y\approx (u(x+\eta)-u(x-\eta))/2\eta
\end{equation}
\noindent This definition leads to an interesting consequence. In the analytic interval  $u(x+\eta)\approx  u(x-\eta)+2\eta \partial_{x}u$  the matching condition at the scale $\eta$ reads:  $\delta_{\eta}u\approx 2\eta \partial_{x}u$ thus defining $\eta$ as a random field. Therefore, to calculate the moments of derivatives, one needs either the  probability density $P(\delta_{\eta}u,\eta)$ or the relation coupling $\eta$ and $\delta_{\eta}u$. The dissipation anomaly enables one to calculate this relation. Here we present the main steps of derivation of Refs. [9], [24], [18].

\noindent We are interested in the Navier-Stokes dynamics of incompressible
fluids, for which the energy balance equation (with the density
$\rho=1$) is written as
$$\frac{1}{2}\frac{\partial
u^{2}}{\partial t}+\frac{1}{2}{\bf u}\cdot \nabla u^{2} =-\nabla
p\cdot {\bf u}+ \nu {\bf u}\cdot \frac{\partial ^{2} {\bf
u}}{\partial x_{i}^{2}},$$.

The differential equation  for the scalar product ${\bf
u(x+\frac{y}{2})\cdot u(x-\frac{y}{2})\equiv u(+)\cdot u(-)}$ can
be written as
\begin{eqnarray}
\frac{\partial {\bf u(+)\cdot u(-)}}{\partial t}+{\bf
u(+)\cdot\frac{\partial}{\partial  x_{+}}u(+)\cdot u(-)}+{\bf
u(-)\cdot\frac{\partial}{\partial x_{-}}u(-)\cdot u(+)}=\nonumber
\\-\frac{\partial p(+)}{\partial x_{+,i}}u_{i}(-)-\frac{\partial
p(-)}{\partial x_{-,i}}u_{i}(+)+ \nu[{\bf u}(-)\cdot
\frac{\partial^{2}}{\partial x_{+,j}^{2}}{\bf u}(+)+{\bf u}(+)\cdot
\frac{\partial^{2}}{\partial x_{-,j}^{2}}{\bf u}(-)].
\end{eqnarray}
It is clear that in the limit $y\rightarrow 0$, for which ${\bf
x}_{\pm}\rightarrow {\bf x}$, this equation gives the energy
balance.  In the limit $y\rightarrow 0$,  the equation (12)
has two kinds of terms:  the regular ones which disappear by the virtue of the energy balance  and a few singular terms balancing each other. The calculation presented in detail in Ref. [18] leads to the equation (See also Duchon and Robert [23], Eyink [24]):
\begin{equation}
lim_{y\rightarrow 0}[- \frac{\partial}{\partial y_{i}}\delta
u_{i}|{\bf \delta_{y} u}|^{2}+\frac{1}{2}(\frac{\partial}{\partial
x_{+,i}}u_{i}(+)u_{j}(-)^{2}+\frac{\partial}{\partial
x_{-,i}}u_{i}(-)u_{j}(+)^{2})=-2{\bf \delta_{y} u \cdot \delta_{y}
a}],
\end{equation}
where ${\bf a}=-\nabla p+\nu \nabla^{2}{\bf u}$ is the lagrangian
acceleration. The above equation, not involving time derivatives,  is exact at each point in the flow at each instant of time.  Choosing the
displacement vector along one of the coordinate axes and averaging, one obtains [24]:
$$\frac{\partial }{\partial y} \overline{\delta u |{\bf \delta
u}|^{2}}=8\nu \overline{\delta u_{i}\frac{\partial^{2}}{\partial
y^{2}}\delta
u_{i}}=2\nu\overline{(\delta_{y}u_{i})\partial_{x}^{2}(\delta_{y}u_{i}})=-\frac{4}{3}{\cal
E},$$ where $\delta_{y} u={\bf \delta_{y} u} \cdot {{\bf y}/y}$. The
pressure terms in and the second contribution 
 disappeared by the averaging procedure. 

\noindent {\bf Dissipation anomaly as a closure.}  The relations  (7)-(8)  are  valid for all magnitudes of the displacement vector $r$ including  $r\rightarrow\eta$.
Substituting the expression for the dissipation anomaly (13)  into the right-side of  (7) gives:

 \begin{eqnarray}
 \frac{\partial S_{2n,0}}{\partial r}+\frac{d-1}{r}S_{2n,0}-\frac{(d-1)(2n-1)}{r}S_{2n-2,2}=-(2n-1)<\delta (\partial_{x}p) u^{2n-2}>=\nonumber \\
\frac{ 3(2n-1)}{4}<[\frac{(\partial (\delta u)^{3}}{\partial r}+(d-1)\frac{\partial \delta_{r}u(\delta_{r} v)^{2}}{\partial r}](\delta_{r}u)^{2n-3}>
 \end{eqnarray}
\noindent The  velocity-kinetic energy product, which appears in equation (13)  has not been included in this relation as small. Using the gaussian boundary conditions implying 
$(2n-1)S_{2n-2,2}=(1+\xi_{2}/2)S_{2n,0}$ this expression can be somewhat simplified.  The full closure is achieved if we express the second contribution to the right side of (14) in terms of $\delta_{r}u$. At this point we cannot do it rigorously but it is plausable to assume  that it is  $O((\delta_{r}u)^{3})$ leading to a generic relation (10).  
  A simple closure  based on the Bernoulli-like relation $\delta_{r}(\partial_{x}p)=O(\partial_{r}(\delta u)^{2})$ leading  to the one-parametrical expression for the exponents $\xi_{n}$ in a close agreement with available experimental data has been proposed in Refs. [12]., [16].\\

\section{Dissipation structures. PDF of dissipation scales.}
 
 \noindent On the dissipation scale $\eta$ all contributions from pressure, advection and dissipation terms are of the same order and the relation for the  dissipation anomaly allows the estimate [18]:

\begin{equation}
\eta\delta_{\eta}u\approx \nu
\end{equation}
 
 \noindent which means that the dissipative structures correspond to the local magnitude of the Reynolds number 
 \begin{equation}
 Re_{\eta}=\eta\delta_{\eta}u/\nu=O(1)
 \end{equation}
This defines $\eta$ as a random field. We would like to stress that the relation (15) has  been obtained  by balancing various  contributions to the locally exact expression for the dissipation anomaly and by establishing the relation between the dissipation scale and velocity increment, this formula enables one to evaluate spatial derivatives and  compute various correlation functions. 
\noindent If  the displacement $y$ is in  the analytic range, then  $u(x+y)-u(x)\approx \partial_{x}u(x)y$ and extrapolating $y\rightarrow \eta$ where $\eta$ is the scale where the analytic and singular ranges overlap, we obtain [18]:
\begin{equation}
\partial_{x}u\approx \delta_{\eta}u/\eta\approx  (\delta_{\eta}u)^{2}/\nu
\end{equation} 
\noindent  acceleration 

\begin{equation}
a\approx \frac{\delta_{\eta} u}{\tau}\approx \frac{(\delta_{\eta} u)^{2}}{\eta}\approx \frac{(\delta_{\eta} u)^{3}}{\nu}
\end{equation}

\noindent and the dissipation rate

\begin{equation}
{\cal E}=\nu {\bf \delta_{\eta} u}\nabla^{2}{\bf \delta_{\eta} u}\approx (\delta_{\eta} u)^{4}/\nu
\end{equation}

\noindent It follows from (17) (See also Ref. [18])  that for  each moment $S_{n,m}$ one can define a "dissipation scale $\eta_{n+m}$ separating analytic and singular ranges and   

\begin{equation}
\eta_{n}\approx LRe^{\frac{1}{\xi_{n}-\xi_{n+1}-1}}
\end{equation}

Using the relations (17)- (20), we can
develop the multi-scaling algebra. For example,
\begin{equation}
\overline{a^{2n}}\approx
(\frac{Re}{u_{rms}L})^{2n}S_{6n}(\eta_{6n})\propto
(\frac{Re}{u_{rms}L})^{2n}\eta_{6n}^{\xi_{6n}}\approx
(\frac{u_{rms}^{2}}{L})^{2n}Re^{a_{2n}},
\end{equation}
with $a_{2n}=2n+\frac{\xi_{6n}}{\xi_{6n}-\xi_{6n+1}-1}$. With
$\xi_{6}=2$ and $\xi_{7}=7/3$, we recover Yaglom's result [24]
$\overline{a^{2}}\approx \frac{u_{rms}^{\frac{9}{2}}}{\sqrt{\nu}}$.
If evaluating the exponents,  one uses in (21) the anomalous exponents $\xi_{6}\approx 1.77$ and $\xi_{7}\approx 1.99$ from Ref.[12],  then $a_{2}\approx 0.55$ meaning that the intermittency correction is $\approx 0.05$.  Recent
experiments by Reynolds et al.\ [25] were in a close agreement with (21).  
A similar formula  has  recently been 
obtained from the multi-fractal formalism by Biferale et al. [26].  Formula (21) shows that the second moment of
Lagrangian acceleration is expressed in terms of the sixth-order
structure function evaluated on its dissipation scale $\eta_{6}$. To
extract information about the fourth moment $\overline{a^{4}}$, we
have to have accurate data on $S_{12}(\eta_{12})$ which, in
high-Reynolds-number flows, is very difficult to obtain from  both physical and, especially, numerical experiments..

The moments  of velocity derivatives  evaluated easily. In
accord with (17) and (20):
\begin{equation}
\overline{(\partial_{x} u)^{2n}}\approx
\overline{(\frac{\delta_{\eta} u}{\eta})^{2n}}\approx
\overline{(\frac{(\delta_{\eta} u)^{2}}{\nu})^{2n}}\approx Re^{2n}S_{4n}(\eta_{4n})\approx 
Re^{d_{2n}},
\end{equation}
where $d_{2n}=2n+\frac{\xi_{4n}}{\xi_{4n}-\xi_{4n+1}-1}$.

\noindent  {\bf Moments of the dissipation rate. The exponent $\mu_{2}$.} Using (19) and (20),  the moments of the dissipation rate are calculated readily: $\overline{\cal E}^{n}\approx Re^{\mu_{n}}$ where $\mu_{n}=n+\xi_{4n}/(\xi_{4n}-\xi_{4n+1}-1)$.  Taking  in accord wit Refs. [[11],[12] 
$\xi_{n}\approx 0.383n/(1+0.05n)$, gives
$$\mu_{2}\approx 0.16$$ and,  since according  to (20)   $\eta_{8}\approx Re^{-0.84}$, we derive 
$\overline{{\cal E}^{2}}\approx \eta_{8}^{-0.19}$.  This result is based on the relation (19) telling us that 
the mean of the square dissipation rate must be evaluated in terms of finite difference (velocity increment) over the region of space having the linear dimension $\eta_{8}$ which is substantially smaller that the Kolmogorov scale $O(Re^{-3/4})$. Extrapolating this relation into an inertial range gives

$$\overline{{\cal E}(x){\cal E}(x+r)}\propto r^{-\mu}$$

\noindent with $\mu\approx 0.19$. The same expression, calculated on the Kolmogorov scale gives 
$\mu\approx 0.21$.

\noindent  It is interesting that the exponents $\mu_{n}$ for $n<1$ are negative. This prediction is yet to be tested experimentally.
Evaluation of the high-order moments of the dissipation rate involving the correlation functions $S_{4n}(\eta_{4n})$  in their respective analytic intervals requires very high resolution of the velocity field and is highly  problematic.  

\noindent {\bf Dissipation scales.} 
According to its definition,  the dissipation scale is a linear dimension of a structure defined by the local value of the Reynolds number $Re_{\eta}=\eta \delta_{\eta}u/\nu=O(1)$.  This introduces a random field. The probability density $Q(\eta,Re,Re_{\eta})$ is found by fixing the displacement $r=\eta$ and counting the  events with $\eta\delta_{r}u/\nu=Re_{\eta}$ keeping the  global and local  Reynolds numbers $Re=u_{rms}L/\nu$ and $Re_{\eta}$  fixed as parameters.  
\noindent Defining $\eta$ as a scale at which the analytic and singular parts of velocity field overlap, 
 naturally leads to the matching condition:  $P(\delta_{\eta}u/\eta^{a},\eta)=P_{D}(\delta_{\eta}u,\eta)$
 where $P_{D}$ is the PDF of $\delta_{r}u$ for the displacement values from the "dissipation" range $r\leq \eta$ where the expression (3) does not work.  On the matching scale however,  the relation (3) is correct.
Thus, the probability density $Q(\eta,Re)$ can be found from $P(\delta_{\eta}u|Re_{\eta}\approx 1)$  where 
$Re_{\eta}=\eta\delta_{\eta}u/\nu$.  From formula  (1): 

\begin{equation}
P(\delta_{\eta} u)\equiv P(u_{\eta})=\frac{1}{u_{\eta}}\int_{-i\infty}^{i\infty} A(n)\nu^{\xi(n)}u_{\eta}^{-\xi(n)-n}dn
\end{equation}
\noindent and  for the scaling exponents given by (3), the result is derived readily. Introducing the large-scale Reynolds number 
$\nu=1/Re$   and taking into account that $Re u_{\eta}\approx 1/\eta$ gives  for the probability density $Q(\eta)$:

\begin{equation}
Q(\eta,Re)=\frac{1}{\eta}\int_{-\infty}^{\infty}e^{-x^{2}}dx\int_{-\infty}^{\infty}dn
 e^{in \ln(\eta^{a+1}\sqrt{2}xRe)-bn^{2}\ln\eta}
\end{equation}
\noindent giving:

\begin{equation}
Q(\eta,Re)=\frac{1}{\eta\sqrt{4b\ln\eta}}\int_{-\infty}^{\infty}
e^{-x^{2}}dx e^{-\frac{\ln^{2}(\eta^{a+1}\sqrt{2}xRe)}{4b\ln \eta}}
\end{equation}

 \noindent The probability density $Q(\eta,Re)$ is plotted on Fig.4a  for $Re=1/\nu=0.3$ and the curves for $Re=0.1;~0.3$ are compared on Fig.4b. \\
 
 \noindent PDF of velocity derivative is computed from (23) combined with (17). The result is:
 
 $$P(u')=\frac{2}{u'}\int_{-i\infty}^{i\infty} A(n)\nu^{\frac{3\xi(n)}{2}}u'^{-\frac{3\xi(n)}{2}-\frac{n}{2}}dn$$

\noindent  The integral is evaluated using the procedure of Section 2.  
\noindent It is interesting that for a "normal" case $\xi_{n}=n/3$  we have:

$$P(u')\propto e^{(\nu u'^{2})^{\frac{2}{3}}}$$

\noindent This relation,  which  has been obtained by Benzi et al [27] from the multi-fractal formalism,  cannot be  a consequence of the Kolmogorov  (K41)  theory.   Indeed,  unlike  the expression (17)  based on the idea of the fluctuating u.v. cut-off,  the K41  dissipation scale $\eta_{K}=const$ is a number not related to $\delta_{\eta}u$ .  Thus,  since the fluctuating dissipation scale is not compatible with K41 and "normal scaling",  this PDF  is not realizable.  If for consistency with K41, one uses the constant dissipation scale, then the PDF is simply the one of $\delta_{\eta_{K}}u$ which, in accord with calculation of Sec 1,  is  a Gaussian. 

\noindent The PDF of the dissipation rate can be easily calculated from the integral  (23) with $u_{\eta}\approx ({\cal E}\nu)^{\frac{1}{4}}$. The resulting expression gives $P_{e}({\cal E},Re)$ 
with the broader tails than  those of Kolmogorov's Log-normal probability density.

\section{Discussion and conclusions.}

\noindent The theory presented in this paper is based on two principle assumptions : ~1. The existence of a   length-scale $L$  such  that for $r=L$ the odd-order moments  $S_{2n+1}(L)\approx 0$.  At this  scale  the energy flux toward small scales sets up and the PDF $P(\delta_{r}u,r=L)$ is  close to the Gaussian. This statement ,  consistent with the Navier-Stokes equations, has been tested in both physical and numerical experiments [19]-[20].  The Gaussian PDF $P(\delta_{L}u,L)$  is not unlike a large-scale boundary condition needed to solve the differential equations (7)-(8) for the moments $S_{n,m}$.  ~ 2.~The expression 
(3) for the first two terms of the Taylor expansion of the function $\xi_{n}$,   consistent with the Holder inequality, is a good approximation to the exponents of the  first 10-15 moments of velocity increments . The existence of a small parameter in turbulence theory  is highly problematic and numerical smallness of  deviations  from the linear expression for not too large moment numbers $n$ in (3)  was   helpful for the theory developed above. The relations  for the dissipation structures and "dissipation"  scales $\eta$ coming  from the order-of-magnitude balancing  of the terms in the exact equations for the dissipation anomaly   are well justified. 

\noindent Using  these assumptions we have shown that the calculated PDF of velocity differences strongly deviates from the Log-normal distribution, first obtained by Kolmogorov in 1962, which  even today is widely used in the literature. This difference stems   from contributions  of the amplitudes $A(n)$ to the integral (1). The amplitudes  $A(n)$ are fixed by the large-scale boundary conditions,  leading to  a natural  conclusion: the small-scale dynamics are strongly coupled to the large-scale phenomena. This  may be  a reason for   a serious reexamination of  the very concept of the turbulence energy cascade which, within the framework of the present development,    seem  neither possible nor needed.   
An accurate experimental and numerical comparison of the measured and Log-normal PDFs of velocity increments may be extremely important. 

\noindent It has been shown  [18] ( in a different way this is also  an element of the multi-fractal theory)   that  the  scales $\eta$ form  a random field not necessarily related to  the energy dissipation scales but rather to the linear dimensions of  various dissipation structures defined by the local value of the Reynolds number $Re_{\eta}=O(1)$.
Some of these  structures are responsible for the energy and  the second-order moment $S_{2,0}$ dissipation, while others, more powerful,  ~ for the dynamics of the higher -order moments. Thus, the scale $\eta$ must be perceived as a dynamic cut-off separating analytic and singular components  of the velocity field.   
\noindent The probability density $Q(\eta,Re)$, calculated in Section 5 is an interesting and easily  measurable quantity. 
A note of caution is in order:  to make reliable calculations or measurements  of the moments involving 
spatial derivatives of velocity field,  one has to have a field resolved well enough to exhibit at least a 
fraction of the analytic range of the corresponding moments of velocity increment. For example, according to (18), the second order  moment of Lagrangian acceleration is proportional to the sixth order structure function calculated on the scale $\eta_{6}$, while the fourth order one is expressed  through $S_{12}(\eta_{12})$. In the high Reynolds number flows, the measurements of the twelveth  order structure function including analytic range where $S_{12}\propto r^{12}$ do not exist. The situation with the moments of dissipation rate is even worse:  
fourth- order moment  $\overline{{\cal E}^{4}}$ is  related to $S_{16}(\eta_{16})$.  An interesting possibility is being explored by Schumacher [28 ]  running very large ($1024^{3}$) numerical simulations at reasonably low Reynolds numbers $R_{\lambda}\approx 10-60$. Analyzing the probability density of the dissipation scales Schumacher and Sreenivasan [29] obtained $Q(\eta,Re)$ very similar to one shown on Fig. 3. 

\noindent The results presented here were obtained from analysis,  both theoretical  and numerical, of the dynamic equations. No multi-fractal assumptions have been made.   Still, it is interesting to compare the two approaches. In its present form, the multi-fractal (MF) theory consists of two parts [1]. The first one based on an idea of fractal dimension,  attempts to explain the origin of anomalous scaling by assuming 

\begin{equation}
S_{p}(r)=(\frac{r}{L})^{\xi_{p}}=\int d\mu(h)(\frac{r}{L})^{ph+3-D(h)}
\end{equation}
\noindent The normalised structure functions $S_{p}=(\frac{\delta_{r}u}{u_{rms}})^{p}$. The $O((r/L)^{ph})$ term comes from the multi-fractal assumption 
\begin{equation}
\delta_{r}u=(r/L)^{h}
\end{equation}

\noindent defined on a set of fractal dimension $D(h)$ where $h$ is a value of the scaling exponent from the interval  $h_{min}\geq h\geq h_{max}$ and $(r/L)^{3-D(h)}$ is the probability of being in the interval $r$ in a volume of dimension $3-D(h)$.   Neglecting the Logs in the steepest descent evaluation of the integral (26),  gives  a relation between the fractal dimension and the exponents $\xi_{n}$ (For the review see Frisch [1]).

\noindent  {\bf  The multifractal PDF  and Mellin transform.} Let us establish  possible relations between the two theories.  Using the gaussian expression for the amplitudes, we have:

\begin{equation}
S_{2p}(r)=(2p-1)!!(\frac{r}{L})^{\xi_{2p}}=(2p-1)!!\int d\mu(h)(\frac{r}{L})^{2ph+3-D(h)}
\end{equation}

\noindent Substituting this into (1) and repeating the calculations of the section 2 gives ($L=1$):

$$P(u,r)=\frac{1}{2}\int e^{-x^{2}}dx \int_{-i\infty}^{i\infty}dn \int
\frac{d\mu(h)}{dh}dhe^{-n(\ln\frac{u}{\sqrt{2}x}-h\ln r)}e^{(3-D(h))\ln r}$$

\noindent Integration over $n$ gives the  delta -function   $\delta(h\ln r-\ln \frac{u}{\sqrt{2}x})$ ($u/x\propto r^{h}$ )  with the finial result:

$$P(u,r)=\frac{1}{2u \ln r }\int  dx e^{-x^{2}}exp (3-D(\frac{\ln \frac{u}{\sqrt{2}{x}}}{|\ln r|}))
\frac{d\mu(h_{*})}{dh}$$

\noindent where $h_{*}=\frac{\ln \frac{u}{\sqrt{2}{x}}}{|\ln r|}$.
Restricitng ourselves by the relation (3) for the exponents and comparing this formula  with the PDF  (5) gives (neglecting the $\mu$-factor ):
\begin{equation}
[3-D(\frac{\ln \frac{u}{\sqrt{2}{x}}}{|\ln r|})]\ln r = -\frac{ (\ln \frac{u}{r^{a}\sqrt{2}x})^{2}}{4b|\ln r|}
\end{equation}

\noindent  No  steepest descent approximation has been used  in deriving this relation. If in accord with the MF theory,  we set  the amplitudes $A(n)=1$ and $\sqrt{2}x=1$  in (29) and taking into account that $h=\ln u/\ln r$, the expression (29) gives:

$$3-D(h)=\frac{ (\ln \frac{u}{r^{a}})^{2}}{4b(\ln r)^{2}} =\frac{(h-a)^{2}}{4b}$$

\noindent  We remind the reader that in accord with Ref. [11],[12], $a\approx 0.383$ and $b\approx 0.0166$. This derivation which did not involve the steepest descent evaluation of the integral does not have  a  "Log-problem",  discussed in Ref. [30].   The experimental measurements of 
Cramer function $f(\alpha)=D(h)+2$ for $h=\alpha/3$ by Meneveau and Sreenivasan [31] (See also Ref.[1]) are in an extremely close agreement with this expression. The quantitative  differences are: 
the maximum of the calculated  curve  (with $\alpha=3h$) is at $\alpha=1.15$ (instead of $\alpha=1$ of Ref. [31]) and $f(\alpha)=0$ at $\alpha_{1}=0.369$ and $\alpha_{2}=1.92$ compared with $\alpha_{1}\approx 0.5$ and 
$\alpha_{2}\approx 1.8$ of Ref.[31]. The small deviations  come  from  the difference between  $h=\alpha/3$
used for analysis of experimental data and  the theoretically obtained $h=a=0.383$. This difference decreases if the coordinates are rescaled by factor $0.383/0.333$.

\noindent  The second part of the MF theory,  dealing  with the small-scale properties of turbulence, is based on the relation (Paladin and Vulpiani [32] )
\begin{equation}
\frac{\eta}{L}\approx Re^{-\frac{1}{1+h}}
\end{equation}

\noindent  obtained by combining the MF assumption 
(27) and the outcome of the balance of the advective and viscous contributions to the Navier-Stokes equations. All  small-scale results derived using the MF formalism are numerically indistinguishable 
from the ones obtained both above and in the Ref. [18]. To illustrate this point,  we present an  alternative derivation of  the moments of velocity derivative  due to Polyakov [33]. 

\noindent {\bf Polyakov's derivation}. The probability density of velocity difference in the inertial range is:

\begin{equation}
P(\delta_{r}u,r)=<\delta(\delta_{r}u-[u(x+r)-u(x)])>
\end{equation}
 \noindent and in the dissipation (analytic)  range
 \begin{equation}
 P_{D}(\delta_{r}u,r)=<\delta(\delta_{r}u-u'r)>
 \end{equation}
\noindent Assuming  the two PDFs match at the scale $r=\eta=\nu/\delta_{\eta}u\equiv \nu/u$,  
introduced   by (15), we have:

\begin{equation}
P_{D}(u,\eta)=P(u,\eta)=<\delta(u-u'(x)\nu/u)>=\int dn A(n)\nu^{\xi_{n}}u^{-\xi_{n}-n-1}
\end{equation} 
 
\noindent Multiplying (33) by $u^{2k}$ and integrating  over $u$  gives:

\begin{equation}
\frac{1}{2}\overline{(\nu u')^{k}}=\int dn A(n)\nu^{\xi_{n}}\frac{1}{2k-\xi_{n}-n}
\end{equation}

\noindent  The integral is evaluated at a pole where $\xi_{n(k)}+n=2k$ giving 

\begin{equation}
\overline{(u'(x))^{k}}\propto A(n(k))\nu^{\rho(k)}
\end{equation}

\noindent with 

\begin{equation}
\rho(k)=\xi_{n(k)}-k=k-n(k)
\end{equation}
\noindent which is identical to  the formula (8.76) of Frisch's book [1] obtained using multi-fractal theory. 
Comparing the relations (35) and (22) we find that , on the accepted magnitudes of  exponents $\xi_{n}$,   numerically they  are basically identical for not too large moment numbers $n$.  It is also easy to see that if  in the limit $n\rightarrow\infty$, $\xi_{n}\propto n^{\alpha}$ with $0\leq \alpha\leq 1$, the two relations have the same asymptotics. 

\noindent The only approximation involved in derivations of  both relations  (22) and (35), presented in this paper, is the 
choice of the cut-off $\eta=\nu/u$ instead of $\eta=O(\nu/u)$. In reality, there exist a random field of the dissipation scales described by the probability density $Q(\eta,Re)$ given by (25). Thus, a more accurate calculation of both (22) and (34) must involve averaging over the fluctuating cut- off $\eta$. However, we do not expect this procedure to introduce substantial modifications of the obtained results. 

\noindent  The  theory presented here does not involve any multi-fractal assumptions.  Still, the quantitative (numerical)  agreements between the two approaches hints on the possibility of some qualitative connection.  The essentially dynamic theory developed here couples  the velocity fluctuations at the largest  and  smallest scales. One may speculate that  if a typical structure is basically a strongly convoluted sheet with two $O(L)$  linear dimensions  and  $O(\eta)$ the third one,  then these  structures can loosely be identified with the multi-ifractal sets of the MF theory.  

\section{Acknowledgements.}  I  am  grateful to A. Polyakov for his comments and discussions which influenced the course of this work.  My gratitude is due to J. Schumacher and K.R. Sreenivasan for their comments and for  sharing  with me their numerical and experimental data on the small-scale statistics and to T. Gotoh  and J. Wanderer for his help in preparation of the manuscript.   Helpful conversations with L. Biferale are gratefully acknowledged.

\noindent  

\noindent {\bf References}\\
\noindent 1.\ U. Frisch, Turbulence.  {\it Legacy of A.N. Kolmogorov}, Cambridge University Press, England (1995)\\
\noindent 2. \ K.R. Sreenivasan and R.A. Antonia, Ann.Rev.Fluid Mech. {\bf 29}, 435 (1997); ~S.Y. Chen, B. Dhruva, S. Kurien, K.R. Sreenivasan
and M.A. Taylor, J.\ Fluid.\ Mech. {\bf 533}, 183 (2005).\\
\noindent 3.\  S. Kurien and K.R. Sreenivasan, Phys.~Rev.~E {\bf 64}, 056302 (2001).\\
\noindent 4. \ T. Gotoh, D. Fukuyama and T. Nakano, Phys. Fluids {\bf 14}; 1065 (2002).\\
\noindent 5.\ V.N. Gribov and A.A. Migdal, Soviet Phys. JETP {\bf 28}; 784, (1968).\\
\noindent 6.\ A.M.Polyakov, "Scale invariance of strong interactions and its application to lepton-hadron reactions, in {\it International School of  High Energy Physics in Erevan, 23 November - 4 December 1971}, (Chernogolovka 1972).\\
\noindent 7.\ K. Gawedzki and A. Kupiainen, Phys. Rev. Lett {\bf 75}, 3834 (1995);~ M. Chertkov, G. Falkovich, I. Kolokolov, and  V. Lebedev, Phys.Rev. {\bf E52}, 4924 (1995). \\
\noindent 8.\ G. Falkovich, K. Gawedzki and M. Vergassola, Rev. Mod. Phys. {\bf 73}; 213 (2001).\\
\noindent 9.\ A.M. Polyakov, Phys.\ Rev.\ E {\bf 52}, 6183 (1995).\\
\noindent 10.\ E. Weinan,  K. Khanin, A. Mazel , Y. Sinai, Phys.Rev.Lett. {\bf 78}, 1904 (1997);\\
 \noindent 11.\ V. Yakhot, Phys.\ Rev.\ E {\bf 63}, 026307 (2001).\\
\noindent 12.\ V.Yakhot, J.\ Fluid Mech.\ {\bf 495}, 135 (2003).\\
\noindent  13.\  A.N. Kolmogorov, J. Fluid Mech. {\bf 13}, 82 (1962).\\
\noindent  14.\ See in:  A.S. Monin and A.M. Yaglom, {\it Statistical Fluid
Mechanics, vol.\ 2}, MIT Press, Cambridge, MA (1975).\\
\noindent 15. \ S.A. Orszag,\ ``Statistical theory of turbulence",
in {\it Fluid Dynamics}, Les Houches, 2237 (1973), eds.\
R. Balian and J.L. Peube, Gordon and Breach, New York.\\
\noindent 16.\ T. Gotoh and T. Nakano,   J.\ Stat.\ Phys. {\bf 113}, 855 (2003).\\
\noindent 17.\ V. Yakhot and K.R. Sreenivasan, Physica A {\bf 343}, 147-155, 2004\\
\noindent 18.\ V. Yakhot and K.R. Sreenivasan,  J. Stat. Phys. , 2005 (in press).\\
\noindent 19.\ D. Donzis, K.R.Sreenivasan and V. Yakhot, (in preparation).\\
\noindent 20.\ A. Karpikov, (in preparation).\\
\noindent 21.\ V. Yakhot, unpublished.\\
\noindent 22.\ R.J. Hill, J.\ Fluid Mech.\ {\bf 368}, 317 (2002).\\
\noindent 23.\ J. Duchon and R. Robert, Nonlinearity {\bf 13}, 249 (2000).\\
\noindent 24.\ G.L. Eyink, Nonlinearity {\bf 16}, 137 (2003).\\
\noindent 25.\ A.M. Reynolds, N. Mordant, A.M. Crawford and E. Bodenschatz, New J.\ Phys.\ art.\ no.\ 58 (2005).\\
\noindent 26.\  L. Biferale, G. Boffetta, A. Celani, B.J. Devenish, A. Lanotte, F.Toschi, {\bf nlin.CD/05041} .\\
\noindent 27.\ R. Benzi, L. Biferale, G. Paladin, A. Vulpiani, M.Vergassola, Phys.Rev.Lett. {\bf 67}, 2299 (1991).\\
\noindent 28.\ J.Schumacher, (in preparation).\\
\noindent 29.\ J Schumacher and K.R. Sreenivasan (In preparation).\\
\noindent 30.\ U. Frisch, M. Martins Afonso, A. Mazzino and V. Yakhot, J. Fluid Mech. {\bf 542}, 97 (2005).\\
\noindent 31.\ C. Meneveau and K.R. Sreenivasan, J. Fluid. Mech. {\bf 224}, 429 (1991).\\
 \noindent 32.\  G. Paladin and A. Vulpiani, Phys.Rep. {\bf 156}, 147 (187).\\
\noindent 33.\ A.M. Polyakov, 2005, private communication.
 \end{document}